\documentclass[prl,twocolumn,superscriptaddress]{revtex4-1}
\usepackage{}
\usepackage{amssymb}
\usepackage{amsfonts}
\usepackage{bbm}
\usepackage{mathrsfs}
\usepackage{graphics,graphicx,epsfig,bm,amsmath,amsthm,amssymb}
\usepackage{bm}
\usepackage{bbm}
\usepackage{longtable}
\usepackage{multirow}
\usepackage{array}
\usepackage{color}
\usepackage[usenames,dvipsnames]{xcolor}

\usepackage{float}
\usepackage[a4paper,colorlinks=true,
linkcolor=blue,citecolor=blue,
pdfauthor={ },
pdftitle={ },
pdfsubject={ },
pdfkeywords={ }]{hyperref}

\begin{document}

\title{Experimental Test of\\Heisenberg's Measurement Uncertainty Relation Based on Statistical Distances}

\author{Wenchao Ma}
\affiliation{Hefei National Laboratory for Physical Sciences at the Microscale and Department of Modern Physics, University of Science and Technology of China, Hefei, Anhui 230026, China}

\author{Zhihao Ma}
\affiliation{Department of Mathematics, Shanghai Jiaotong University, Shanghai 200240, China}

\author{Hengyan Wang}
\affiliation{Hefei National Laboratory for Physical Sciences at the Microscale and Department of Modern Physics, University of Science and Technology of China, Hefei, Anhui 230026, China}

\author{Zhihua Chen}
\affiliation{Department of Applied Mathematics, Zhejiang University of Technology, Hangzhou, Zhejiang 310014, China}

\author{Ying Liu}
\affiliation{Hefei National Laboratory for Physical Sciences at the Microscale and Department of Modern Physics, University of Science and Technology of China, Hefei, Anhui 230026, China}

\author{Fei Kong}
\affiliation{Hefei National Laboratory for Physical Sciences at the Microscale and Department of Modern Physics, University of Science and Technology of China, Hefei, Anhui 230026, China}

\author{Zhaokai Li}
\affiliation{Hefei National Laboratory for Physical Sciences at the Microscale and Department of Modern Physics, University of Science and Technology of China, Hefei, Anhui 230026, China}
\affiliation{Synergetic Innovation Center of Quantum Information and Quantum Physics, University of Science and Technology of China, Hefei, Anhui 230026, China}

\author{Xinhua Peng}
\affiliation{Hefei National Laboratory for Physical Sciences at the Microscale and Department of Modern Physics, University of Science and Technology of China, Hefei, Anhui 230026, China}
\affiliation{Synergetic Innovation Center of Quantum Information and Quantum Physics, University of Science and Technology of China, Hefei, Anhui 230026, China}

\author{Mingjun Shi}
\affiliation{Hefei National Laboratory for Physical Sciences at the Microscale and Department of Modern Physics, University of Science and Technology of China, Hefei, Anhui 230026, China}
\affiliation{Synergetic Innovation Center of Quantum Information and Quantum Physics, University of Science and Technology of China, Hefei, Anhui 230026, China}

\author{Fazhan Shi}
\affiliation{Hefei National Laboratory for Physical Sciences at the Microscale and Department of Modern Physics, University of Science and Technology of China, Hefei, Anhui 230026, China}
\affiliation{Synergetic Innovation Center of Quantum Information and Quantum Physics, University of Science and Technology of China, Hefei, Anhui 230026, China}

\author{Shao-Ming Fei}
\affiliation{School of Mathematical Sciences, Capital Normal University, Beijing 100048, China}

\author{Jiangfeng Du}
\email{djf@ustc.edu.cn}
\affiliation{Hefei National Laboratory for Physical Sciences at the Microscale and Department of Modern Physics, University of Science and Technology of China, Hefei, Anhui 230026, China}
\affiliation{Synergetic Innovation Center of Quantum Information and Quantum Physics, University of Science and Technology of China, Hefei, Anhui 230026, China}

\begin{abstract}
Incompatible observables can be approximated by compatible observables in joint measurement or measured sequentially, with constrained accuracy as implied by Heisenberg's original formulation of the uncertainty principle. Recently, Busch, Lahti, and Werner proposed inaccuracy trade-off relations based on statistical distances between probability distributions of measurement outcomes [Phys.~Rev.~Lett. \textbf{111}, 160405 (2013); Phys.~Rev.~A \textbf{89}, 012129 (2014)]. Here we reformulate their theoretical framework, derive an improved relation for qubit measurement, and perform an experimental test on a spin system. The relation reveals that the worst-case inaccuracy is tightly bounded from below by the incompatibility of target observables, and is verified by the experiment employing joint measurement in which two compatible observables designed to approximate two incompatible observables on one qubit are measured simultaneously.
\end{abstract}
\maketitle

The uncertainty principle was first proposed by Heisenberg in the context of a measurement process \cite{Heisenberg}. He conceived of a $\gamma$-ray microscope and pointed out that the measurement of an electron's position $Q$ disturbs the momentum $P$ inevitably, and that the product of the error $\varepsilon (Q)$ and disturbance $\eta (P)$ cannot be arbitrarily small. A qualitative relation was written as $\varepsilon (Q)\eta (P)\sim h$, where $h$ is Planck's constant.
In the ensuing few years, Kennard \cite{Kennard}, Weyl \cite{Weyl}, Robertson \cite{Robertson}, and Schr\"{o}dinger \cite{Schroedinger} derived mathematically rigorous versions, including the famous relation $\sigma (A)\sigma (B) \ge \frac{1}{2}\left| {\left\langle {[A,B]} \right\rangle } \right|$, where the angle brackets represent expectations, the standard deviation $\sigma \left( A \right) = \sqrt {\langle {{A^2}} \rangle  - {\langle A \rangle^2}}$ is called preparation uncertainty, and $[A,B] = AB - BA$.
However, these formal inequalities, together with well-developed entropic uncertainty relations \cite{Hirschman,Beckner,Birula,Deutsch,Kraus,Maassen,Wehner}, do not handle the problem Heisenberg discussed, but refer to the uncertainties intrinsic to quantum states.


The attempts to quantify Heisenberg's original idea have a long history \cite{Busch07}. A dozen years ago, Ozawa proved a universally valid error-disturbance relation \cite{Ozawa03PRA}. Shortly thereafter, Hall and Ozawa independently showed that the inaccuracies of any joint measurement estimating two incompatible observables satisfy similar relations \cite{Ozawa03IJQI,Ozawa04PLA,Hall}. Later, tighter inequalities were derived by Branciard \cite{Branciard13,Branciard14} and Weston \emph{et al.} \cite{Weston}. These relations have been experimentally verified using polarized neutrons \cite{Erhart12, Sulyok13} and photons \cite{Rozema12,Baek13,Weston,Ringbauer,Kaneda}. Nevertheless, the physical validity of the definitions of error and disturbance in Ozawa's relation is in dispute \cite{Lorenzo,Korzekwa,Dressel,Busch04PLA,Werner,Busch13PRL,Busch14PRA}. Information-theoretic definitions for noise and disturbance were introduced and a state-independent tradeoff relation was derived by Buscemi \emph{et al} \cite{Buscemi}, and this relation has been verified by neutron spin qubits \cite{Sulyok15}.

Recently, Busch, Lahti, and Werner (BLW) formulated uncertainty relations dealing with the imprecisions in joint measurements approximating incompatible observables \cite{Busch13PRL,Busch14PRA,Busch14RMP}.
The imprecisions or uncertainties, based on statistical distances between probability distributions of measurement outcomes, are processed into state independent and regarded as figures of merit of joint-measurement devices. Besides, the joint measurement scenario covers the successive measurement scenario with errors and disturbances.
In this work, we reformulate BLW's theoretical framework, derive an improved relation for qubit measurement, and perform an experimental test on a spin system.

Consider a pair of incompatible observables $A,B$. Incompatible means $A,B$ are not jointly measurable, i.e., cannot be measured simultaneously.
For qubit measurements, $A,B$ can be selected as two sharp observables with the spectral projections ${A_ \pm } = (\mathbbm{1} \pm {\bm{a}} \cdot {\bm{\sigma }})/2$ and ${B_ \pm }$ $=(\mathbbm{1} \pm {\bm{b}} \cdot {\bm{\sigma }})/2$, where ${\bm a},{\bm b}$ are unit vectors. If ${\bm{a}}$ and ${\bm{b}}$ are noncollinear, the observables $A,B$ are incompatible.
Approximate measurements of $A,B$ can be performed simultaneously by a device composed of compatible observables $C,D$, each being used as an approximation of $A,B$, respectively.
The observable $C$ consists of positive operators $C_+  = ({c_0}\mathbbm{1} + {\bm{c}} \cdot {\bm{\sigma }})/2$, $C_- = \mathbbm{1} - C_+$, with $\left| {\bm{c}} \right| \le \min \{ {c_0},2 - {c_0}\}  \le 1$ for the positivity. Similarly, the observable $D$ consists of $D_+  = ({d_0}\mathbbm{1} + {\bm{d}} \cdot {\bm{\sigma }})/2$, $D_- = \mathbbm{1} - D_+$, with $\left| {\bm{d}} \right| \le \min \{ {d_0},2 - {d_0}\}  \le 1$.
Compatible or jointly measurable requires that there exists a positive-operator valued measure (POVM) $M$ of which $C,D$ are the marginals, namely, $C_i = \sum\limits_j {{M_{ij}}}$ and $D_j = \sum\limits_i {{M_{ij}}}$, with $i,j = +$ or $-$ (hereinafter the same) and $M_{ij} = ({{C_i}{D_j} + {D_j}{C_i}})/2$. When $c_0=d_0=1$, the criterion of compatibility takes a simple form as \cite{Busch14PRA,Busch86PRD,Yu}
\begin{equation}\label{eq:compatible}
\left| {{\bm{c}} + {\bm{d}}} \right| + \left| {{\bm{c}} - {\bm{d}}} \right| \le 2.
\end{equation}
Compatibility allows $C,D$ to be performed simultaneously, but imposes a strong restriction that prevents $C,D$ approaching $A,B$ freely. The overall deviation of $C,D$ from $A,B$, or the performance of the device, is rooted in probability distributions of measurement outcomes.

\begin{figure}\centering
\includegraphics[width=1\columnwidth]{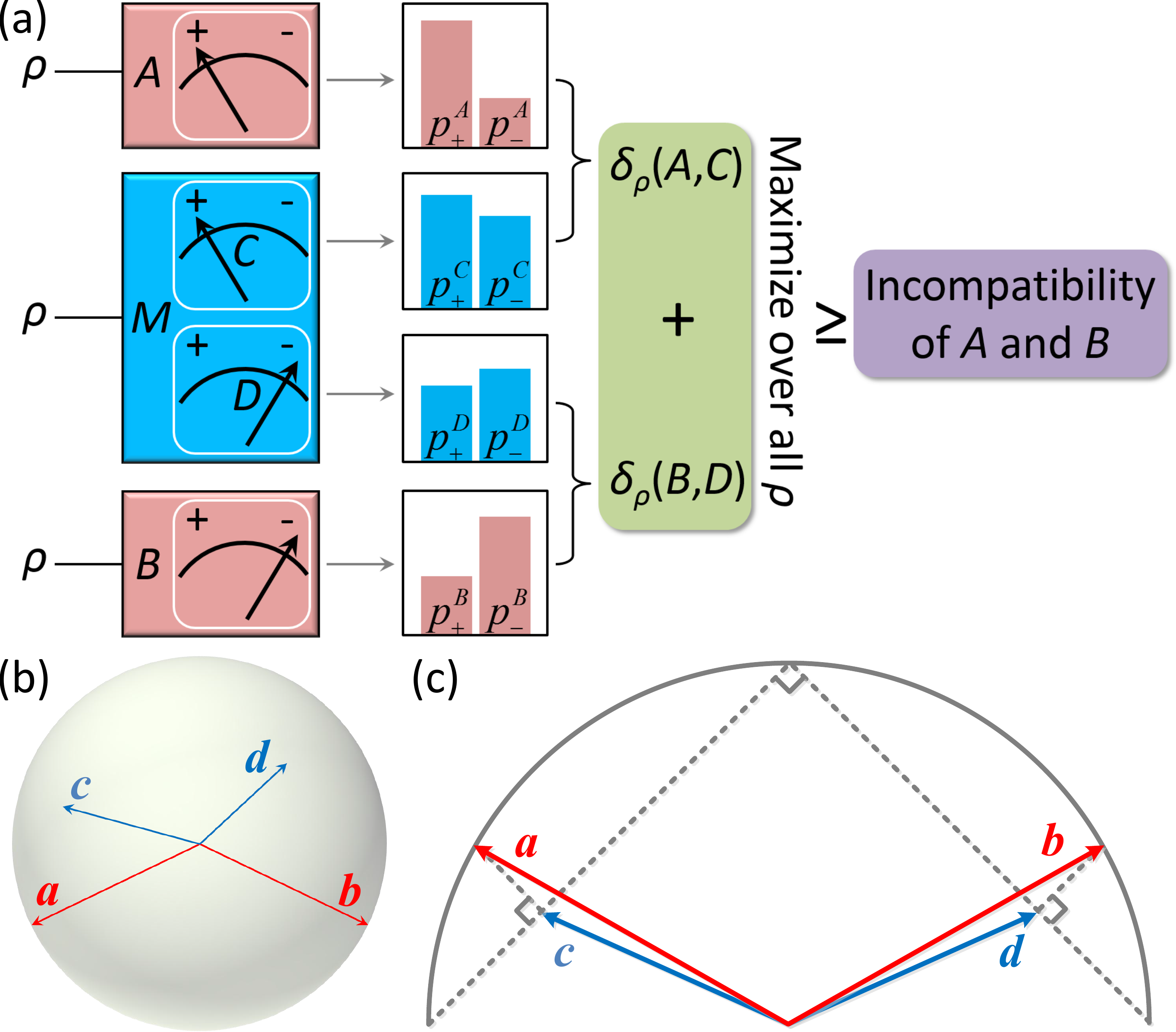}
\caption{(color online). Sketch of the theory.
   (a) Outline of the inaccuracy trade-off relation in Eq.~(\ref{newerror}). Here compatible observables $C,D$ are jointly measured to approximate incompatible observables $A,B$. 
   (b) An example of Bloch vectors ${\bm{a}},{\bm{b}},{\bm{c}},{\bm{d}}$. The lengths of ${\bm{c}},{\bm{d}}$ are usually less than $1$.
   (c) Bloch vectors ${\bm{c}},{\bm{d}}$ of the optimal compatible observables $C,D$ that approximate $A,B$. The dashed chords are legs of an inscribed isosceles right triangle. The vectors $\bm{a}-\bm{c}$ and $\bm{b}-\bm{d}$ are orthogonal to these dashed chords, and the end points of ${\bm{c}},{\bm{d}}$ are the perpendicular feet. The whole pattern is coplanar and symmetric. 
   }
    \label{theory}
\end{figure}

The operators $O_ \pm$ ($O = A,B,C,D$) are called effects, and the probabilities of measurement outcomes are $p_\pm ^O= {\rm{tr}}\left( {\rho {O_ \pm }} \right)$, where $\rho  = ({\mathbbm{1} + {\bm{r}} \cdot {\bm{\sigma }}})/2$ is the density operator of a qubit. We scale the outcomes of these measurements to be $\pm1$, and then the observable $O$ is given as a map $\pm 1 \mapsto {O_ \pm }$.
The statistical difference between the measurements of $A$ and $C$, namely, the inaccuracy or uncertainty, can be quantified as ${\delta_\rho }\left( {A,C} \right) := 2\sum\limits_i {\left| {p_i^A - {\rm{ }}p_i^C} \right|}  = 2\left| {1 - {c_0} + {\bm{r}} \cdot ({\bm{a}} - {\bm{c}})} \right|$ \cite{Busch14PRA}. Likewise, the statistical difference between the measurements of $B$ and $D$ is written as ${\delta_\rho }\left( {B,D} \right) := 2\sum\limits_i {\left| {p_i^B - {\rm{ }}p_i^D} \right|}  = 2\left| {1 - {d_0} + {\bm{r}} \cdot ({\bm{b}} - {\bm{d}})} \right|$.
Since the observables $C,D$ are jointly measured, they face the same quantum state $\rho$ and the combined difference $\Delta_\rho (A,B;C,D) := {\delta_\rho }\left( {A,C} \right) + {\delta_\rho }\left( {B,D} \right)$ is the inaccuracy of the joint measurement for a specific state. The state-dependent quantity $\Delta_\rho (A,B;C,D)$ has a vanishing lower bound for any given $A,B$, and $\rho$. In other words, $\Delta_\rho (A,B;C,D)$ vanishes when minimizing over all $C,D$ for any given $A,B$, and $\rho$ (see Sec.~II.A in Ref.~\cite{sm}). By maximizing $\Delta_\rho(A,B;C,D)$ over all $\rho$, one obtains $\Delta (A,B;C,D) := {\max \limits_\rho }~\Delta _\rho (A,B;C,D)$ as the worst-case inaccuracy of the joint measurement. The state-independent quantity $\Delta(A,B;C,D)$ characterizes the integrated deviation of $C,D$ from $A,B$ and thus is a figure of merit of the measurement device. The inaccuracy trade-off relation is
\begin{equation}\label{newerror}
\begin{aligned}
\Delta(A,B;C,D)  &:=  {\max _\rho }\left[ {{\delta_\rho }\left( {A,C} \right) + {\delta_\rho }\left( {B,D} \right)} \right] \\
\ge \Delta_{\rm{lb}}(A,B) &:= \left| {{\bm{a}} + {\bm{b}}} \right| + \left| {{\bm{a}} - {\bm{b}}} \right| - 2,
\end{aligned}
\end{equation}
as outlined in Fig.~\ref{theory}(a). The lower bound $\Delta_{\rm{lb}}(A,B)$ is the amount of violation of the inequality (\ref{eq:compatible}), and is termed as the degree of incompatibility of $A,B$ \cite{Busch14PRA}. This lower bound is attained when $C,D$ are the best approximations of $A,B$, and such $C,D$ have $c_0=d_0=1$ and ${\bm{c}},{\bm{d}}$ depicted in Fig.~\ref{theory}(c).
Compared with BLW's original relation, the characteristic quantity $\Delta(A,B;C,D)$ possesses an explicit physical meaning as the worst-case inaccuracy and has a delicate lower bound without an extra coefficient (see Sec.~II.B in Ref.~\cite{sm}). Additionally, as a modification of BLW's conceptual framework, the improved approach also applies to any pair of incompatible observables such as position and momentum although the combined inaccuracy may not be in the additive form.

\begin{figure*}\centering
\includegraphics[width=1.6\columnwidth]{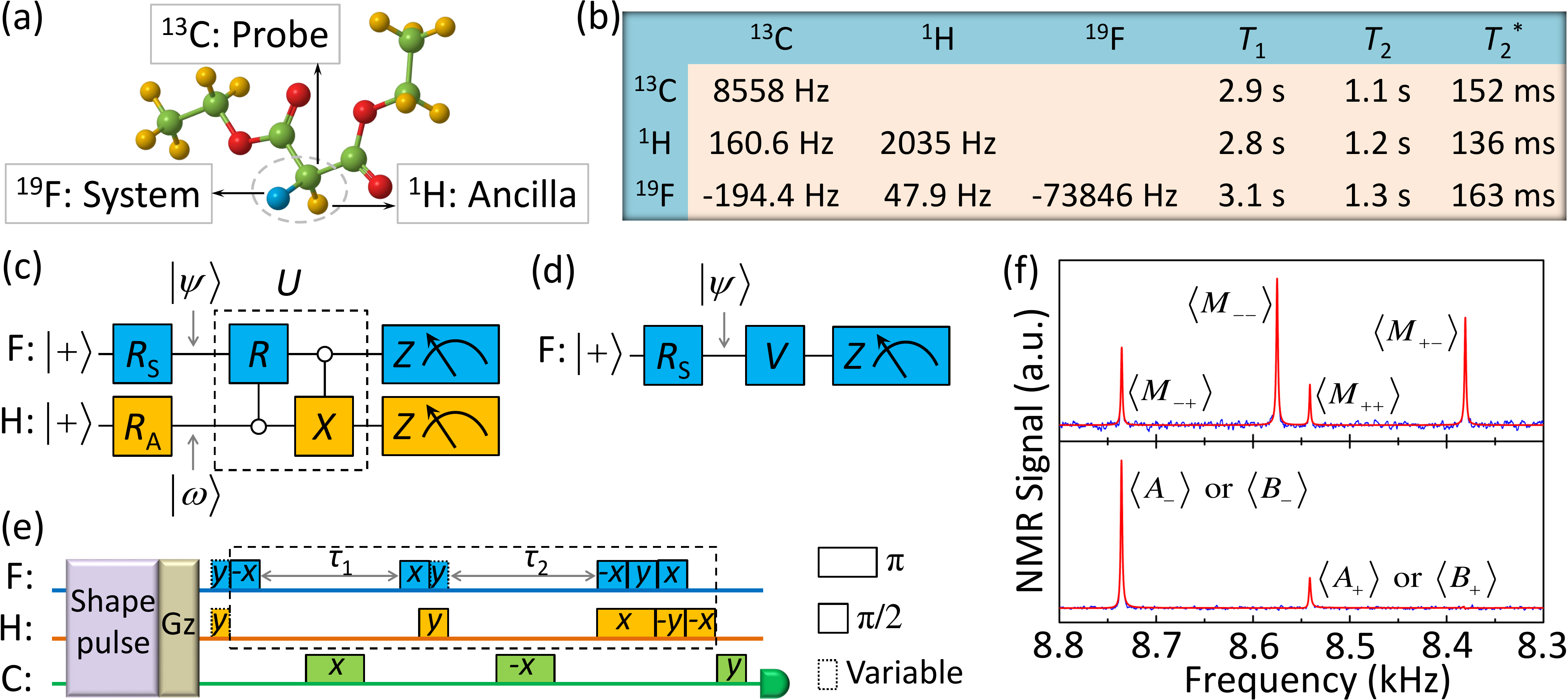}
\caption{(color online). Experimental system and methods.
   (a),(b) Molecular structure and NMR parameters of diethyl fluoromalonate. The chemical shifts and scalar couplings are on and below the diagonal of the table. The relaxation times are in the right half of the table.
   (c) Quantum circuit for the joint measurement of $C,D$.
   (d) Quantum circuit for the measurement of $A$ or $B$.
   (e) Experimental process for the joint measurement of $C,D$. All the square pulses are hard pulses with negligible duration compared to the free evolution time $\tau_1$ and $\tau_2$. The operation $U$ is implemented by the pulse sequence in the dashed frame (see Sec.~III in Ref.~\cite{sm}). The $\pi$ pulses applied on $^{13}$C are for decoupling, and the final $\pi/2$ pulse is for readout.
   (f) $^{13}$C spectra for the joint measurement of $C,D$ via $M$ (upper panel) and the measurement of $A$ or $B$ (lower panel). The jittering blue curves represent the experimental data and the smooth red curves are from Lorentzian fitting.
}
    \label{nmr}
\end{figure*}

The nuclear spin system, operated on a 400 MHz liquid-state nuclear magnetic resonance (NMR) spectrometer, is employed to experimentally demonstrate the relation in Eq.~(\ref{newerror}).
We use diethyl fluoromalonate as the sample dissolved in $^2$H-labeled chloroform at 304 K.
As shown in Fig.~\ref{nmr}(a), the $^{19}$F, $^1$H, and $^{13}$C nuclear spins serve as the system, ancilla, and probe qubits, respectively.
The Hamiltonian of the three-qubit system in the triple-resonance rotating frame is
\begin{equation}\label{hamiltonian}
H = 2\pi \sum\limits_{1 \le k < l \le 3} {{J_{kl}}I_z^kI_z^l},
\end{equation}
with the scalar couplings $J_{kl}$ listed in Fig.~\ref{nmr}(b).

The experiment begins with preparing the pseudopure state (PPS) $\rho_{\rm{pps}} = (1 - \varepsilon )\mathbbm{1}/8 + \varepsilon \left| { +  +  + } \right\rangle \left\langle { +  +  + } \right|$ from the thermal equilibrium state using the line-selective method \cite{PPS,LSpure}. Here $\varepsilon  \approx {10^{ - 5}}$ denotes the polarization and $\mathbbm{1}$ denotes the $8 \times 8$ identity matrix. In this paper, $\left|+ \right\rangle$ and $\left|- \right\rangle$ represent the eigenvectors of the Pauli matrix $Z$.
A shape pulse based on the gradient ascent pulse engineering (GRAPE) algorithm \cite{GRAPE} and a pulsed field gradient are utilized for this step.
After such initialization, a local operation $R_{\rm{S}}$ on the system qubit prepares the state $\left| \psi \right\rangle = R_{\rm{S}} \left| + \right\rangle$ that maximizes $\Delta_\rho$. In our experiment, all involved $C,D$ have $c_0=d_0=1$, and the state $\left| \psi \right\rangle$ is simply determined by $\bm{a},\bm{b},\bm{c},\bm{d}$ (see Section IV in Ref.~\cite{sm}).
If the angle between the vectors $\bm{a}-\bm{c}$ and $\bm{b}-\bm{d}$ is an acute or obtuse angle, the Bloch vector of $\left| \psi \right\rangle$ should be collinear with $\bm{a}+\bm{b}-\bm{c}-\bm{d}$ or $\bm{a}-\bm{b}-\bm{c}+\bm{d}$, respectively. If the angle is a right angle, the Bloch vector should be collinear with either.

For the measurement, different schemes are utilized to implement the observables $A,B$ and $C,D$.
The compatible observables $C,D$ are measured in a joint way. In other words, we measure the joint observable $M$ induced by $C,D$ as previously mentioned. As a POVM with four components, $M$ can be extended to orthogonal projective measurements on a certain two-qubit basis $\left| {{\chi _{ij}}} \right\rangle$ by encompassing the ancilla qubit.
The logic circuit is illustrated in Fig.~\ref{nmr}(c).
A local operation $R_{\rm{A}}$ on the ancilla qubit prepares the state $\left| \omega \right\rangle = R_{\rm{A}} \left| + \right\rangle$.
After that, a global operation $U$ on the two qubits is performed with the controlled gates realized by hard pulses and free evolution.
Finally, the probabilities on $\left| ij \right\rangle$ is measured. The combined effect of $U$ and the measurement on $\left| ij \right\rangle$ amounts to the measurement on $\left| {{\chi _{ij}}} \right\rangle = {U^\dag}\left| ij \right\rangle$, and the POVM on the system qubit is yielded as $M_{ij} = {\rm{tr}}{_{\rm{A}}}\left[ {\left| {{\chi _{ij}}} \right\rangle \left\langle {{\chi _{ij}}} \right|\left( {\mathbbm{1} \otimes \left| \omega  \right\rangle \left\langle \omega  \right|} \right)} \right]$, where $\mathbbm{1}$ denotes the identity operator on the system qubit and ${\rm{tr}}{_{\rm{A}}}$ the partial trace over the ancilla. The explicit form of $M$ is
\begin{equation}\label{M}
\begin{aligned}
 &{M_{ \pm  \pm }} = {\left| {\left\langle { - }
 \mathrel{\left | {\vphantom { -  \omega }}
 \right. \kern-\nulldelimiterspace}
 {\omega } \right\rangle } \right|^{\rm{2}}}\left|  \pm  \right\rangle \left\langle  \pm  \right|, \\
 &{M_{ \pm  \mp }} = {\left| {\left\langle { + }
 \mathrel{\left | {\vphantom { +  \omega }}
 \right. \kern-\nulldelimiterspace}
 {\omega } \right\rangle } \right|^{\rm{2}}}{R^\dag }\left|  \pm  \right\rangle \left\langle  \pm  \right|R. \\
\end{aligned}
\end{equation}
In this way, different $M$ and thus $C,D$ can be implemented by adjusting the operations $R$ and $R_{\rm{A}}$.
As the upper panel of Fig.~\ref{nmr}(f) shows, the statistics of $M$'s four outcomes are displayed as four peaks on the NMR spectrum at once by introducing the $^{13}$C nuclear spin as the probe qubit. The peak areas are proportional to the probabilities on $\left| ij \right\rangle$, and hence the probabilities relating to $M$. Therefore, the two compatible but typically non-commutative observables $C,D$ are simultaneous measured with the outcome distributions derived from that of $M$.

The incompatible observables $A$ and $B$ are measured separately, and the ancilla qubit is not required. The logic circuit is illustrated in Fig.~\ref{nmr}(d). The system qubit in the state $\left| \psi \right\rangle$ is measured on the basis ${V^\dag}\left| i \right\rangle$ constructed as the eigenstates of $A_\pm$ or $B_\pm$. The statistics of measurement outcomes are displayed as two peaks on the spectrum as the lower panel of Fig.~\ref{nmr}(f) shows. By comparing the outcome distributions of $C,D$ with that of $A,B$, one obtains the state-dependent inaccuracy $\Delta_\rho$. For the special preselected state $\left| \psi \right\rangle$, the value of $\Delta_\rho$ is identical to the state-independent inaccuracy $\Delta$.

\begin{figure}\centering
\includegraphics[width=1\columnwidth]{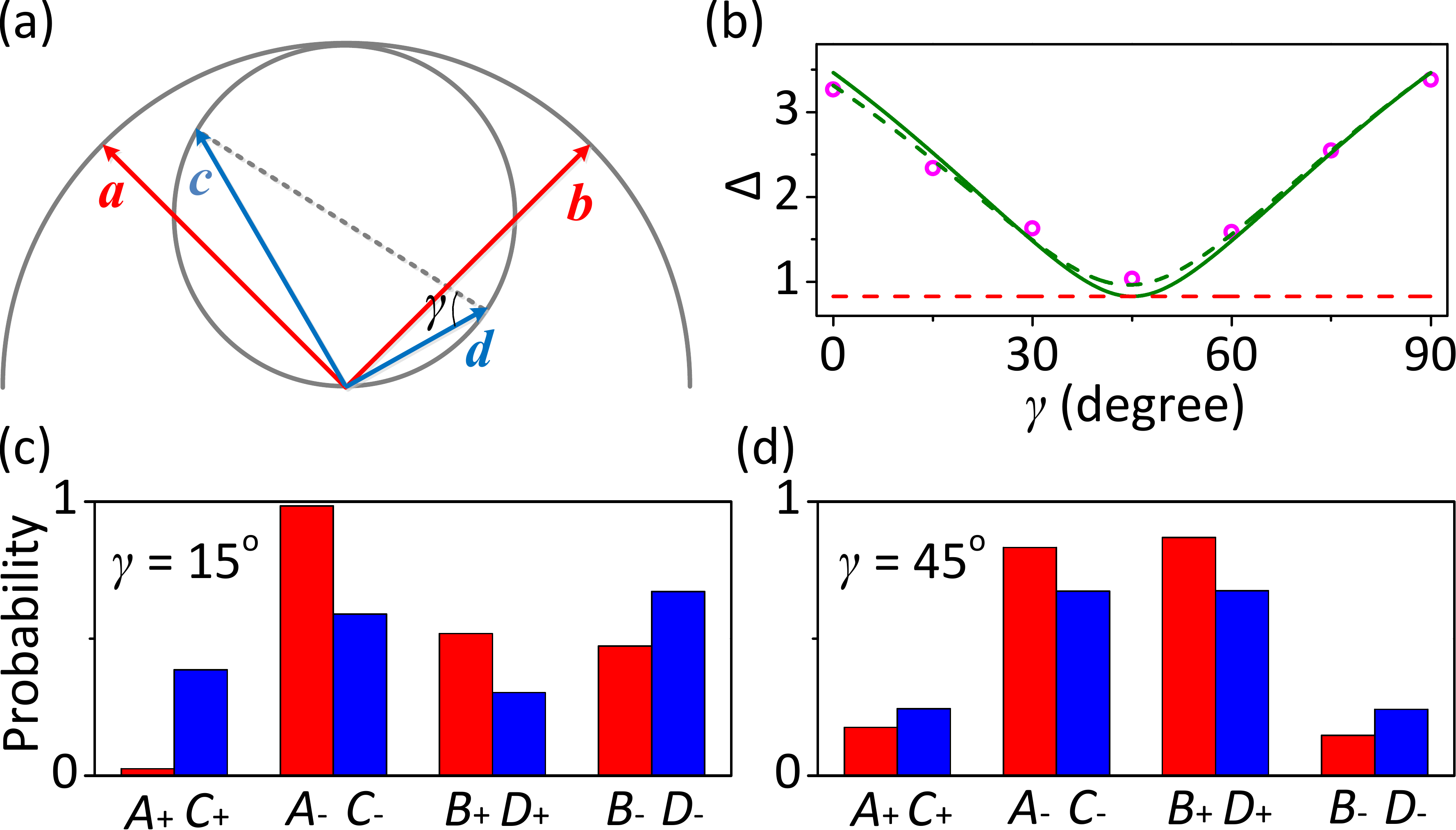}
\caption{(color online). Experimental demonstration of the inaccuracy trade-off relation with maximally incompatible $A,B$.
   (a) Bloch vectors of observables. The whole configuration is coplanar. The vectors ${\bm{a}},{\bm{b}}$ are mutually perpendicular and incline symmetrically about the vertical direction, which is the $z$ direction. The end points of vectors ${\bm{c}},{\bm{d}}$ are the diameter ends of the small circle which is internally tangent to the semicircle and passes through its center. The center of the semicircle is the origin and the tangent point is in the $z$ axis.
   (b) The inaccuracy $\Delta$ as a function of the angle $\gamma  = \arctan \left( {\left| {\bm{c}} \right|/\left| {\bm{d}} \right|} \right)$. The circles, solid curves, and dashed curves represent the experimental, theoretical, and simulated values of $\Delta$, respectively. The numerical simulation takes decoherence into account. The dashed straight line denotes the lower bound $\Delta_{\rm{lb}}$.
   (c),(d) Experimental statistics of measurement outcomes for $\gamma=15^\circ,45^\circ$.
}
    \label{verify1}
\end{figure}

\begin{figure}\centering
\includegraphics[width=1\columnwidth]{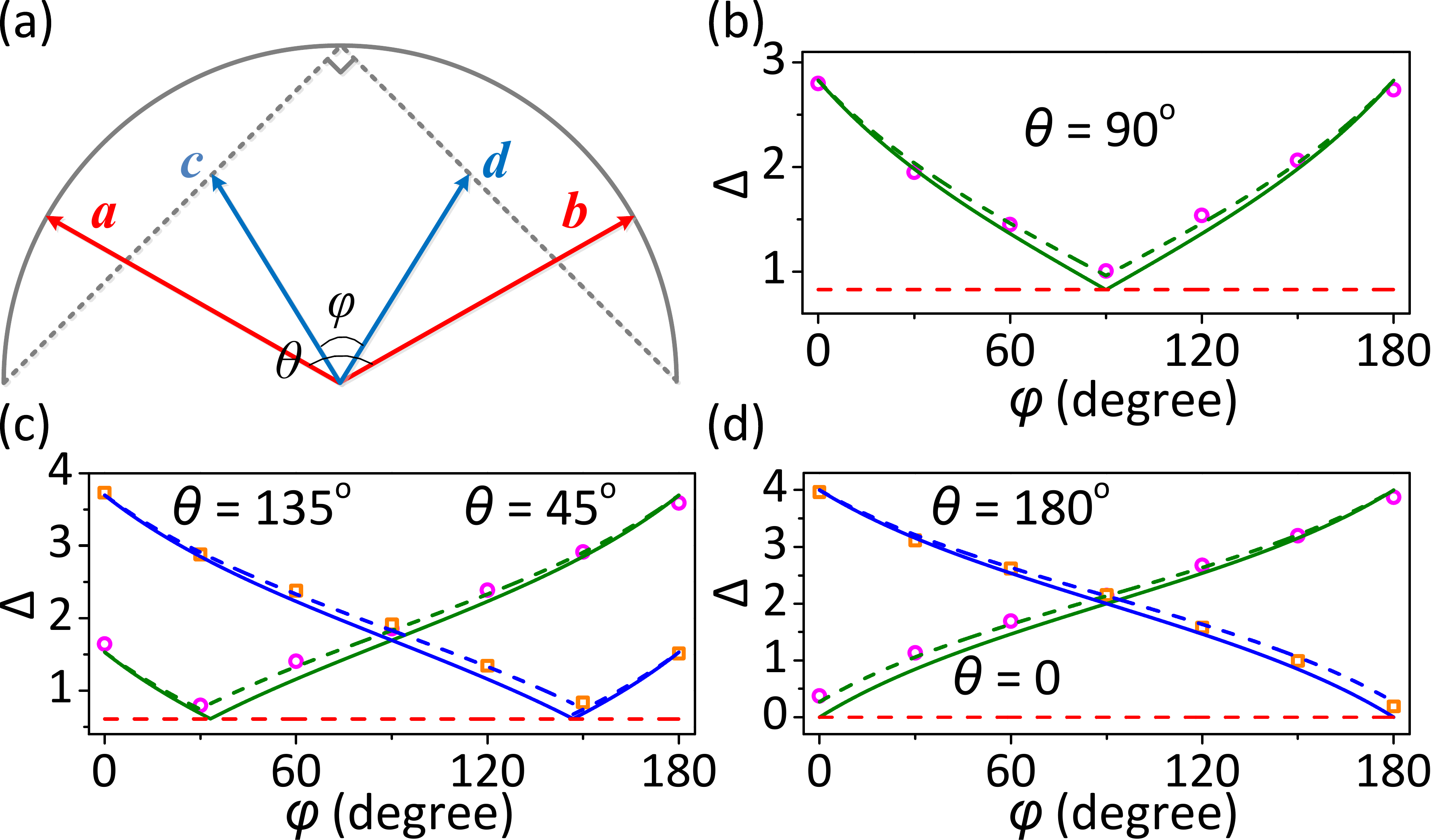}
\caption{(color online). Experimental demonstration of the inaccuracy trade-off relation with several pairs of $A,B$.
   (a) Bloch vectors of observables. The whole configuration is coplanar and symmetric. The dashed chords have the same meaning as that in Fig.~\ref{theory}(c). The angle between the vectors ${\bm{a}},{\bm{b}}$ is $\theta$ and the angle between the vectors ${\bm{c}},{\bm{d}}$ is $\varphi$.
   (b)-(d) The inaccuracy $\Delta$ as a function of the angle $\varphi$ for several pairs of $A,B$ with $\theta=90^\circ$(b), $\theta=45^\circ,135^\circ$(c), and $\theta=0^\circ,180^\circ$(d). The lower bound $\Delta_{\rm{lb}}$ varies with $\theta$, and is the same for $\theta$ and $180^\circ-\theta$. The symbols in these diagrams have the same meanings as that in Fig.~\ref{verify1}(b). 
}
    \label{verify2}
\end{figure}

To experimentally validate the inaccuracy trade-off relation, we select several configurations of $A,B$ and $C,D$ with all the Bloch vectors in the $xz$ plane. Every pair of $\bm{a},\bm{b}$ is symmetric about the $z$ axis, and is expressed as ${\bm{a}} = \left( { - \sin (\theta /2),\cos (\theta /2)} \right),{\bm{b}} = \left( {\sin (\theta /2),\cos (\theta /2)} \right)$, where the former and latter components of a vector correspond to $x$ and $z$, respectively (hereinafter the same). In the case where $\theta=90^\circ$, the vectors $\bm{a},\bm{b}$ are perpendicular to each other and the relevant observables $A,B$ are maximally incompatible. Consider a series of compatible observables $C,D$ with ${\bm{c}} = \left( { - \sin \gamma \cos \gamma ,{{\sin }^2}\gamma } \right),{\bm{d}} = \left( {\sin \gamma \cos \gamma ,{{\cos }^2}\gamma } \right)$ as illustrated in Fig.~\ref{verify1}(a). The inaccuracy $\Delta$ varies with $\gamma$ and is bounded by $\Delta_{\rm{lb}}=2\left( {\sqrt 2  - 1} \right)\approx 0.83$ as shown in Fig.~\ref{verify1}(b). When $\gamma=45^\circ$, the corresponding $C,D$ are the optimal approximations to $A,B$ and the inaccuracy $\Delta$ reaches its lower bound $\Delta_{\rm{lb}}$. The statistics of measurement outcomes for two instances where $\gamma=15^\circ,45^\circ$ are listed in Figs.~\ref{verify1}(c) and \ref{verify1}(d), which visually support the superiority of the optimal $C,D$.
Another series of $C,D$ with ${\bm{c}} = \left( { - 1/\left[ {1 + \cot (\varphi /2)} \right],1/\left[ {1 + \tan (\varphi /2)} \right]} \right),{\bm{d}} = \left( {1/\left[ {1 + \cot (\varphi /2)} \right],1/\left[ {1 + \tan (\varphi /2)} \right]} \right)$ are illustrated in Fig.~\ref{verify2}(a). For $A,B$ with $\theta=90^\circ$, the optimal $C,D$ are reached when $\varphi=90^\circ$ as shown in Fig.~\ref{verify2}(b). This series of $C,D$ are also used to approximate other $A,B$ which are less incompatible.
For $\theta=45^\circ$ or $135^\circ$, the inaccuracy $\Delta$ is bounded by $\Delta_{\rm{lb}}=2/\sqrt {2 - \sqrt 2 }  - 2 \approx 0.61$ as illustrated in Fig.~\ref{verify2}(c). In the limit where $\theta=0^\circ$ or $180^\circ$, $A,B$ reduce to compatible with a vanishing $\Delta_{\rm{lb}}$ as shown in Fig.~\ref{verify2}(d), and the optimal $C,D$ are just $A,B$. All the experimental results confirm the inaccuracy trade-off relation in Eq.~(\ref{newerror}).

%

In conclusion, we have reformulated BLW's theoretical framework, derived an improved relation for qubit measurement, and performed an experimental test using the NMR technique.
We show that as a figure of merit of the measurement device, the worst-case inaccuracy is tightly bounded by the incompatibility of target observables in the qubit case. In the experiment, the device is simulated by joint measurement which measures two compatible observables simultaneously.
Our work represents an advance in quantitatively understanding and experimental verification of Heisenberg's uncertainty principle, and could have implications for the area of quantum information technology.

The authors thank Hui Zhou for helpful discussions. This work was supported by the 973 Program (Grant No.~2013CB921800), the National Natural Science Foundation of China (Grants No.~11227901, No.~31470835, No.~11371247, No.~11201427, No.~11571313, No.~11575173, No.~11575171, and No.~11275131), the China Postdoctoral Science Foundation, the CAS (Grant No.~XDB01030400), and the Fundamental Research Funds for the Central Universities (WK2340000064).
W.~M., Z.~M., and H.~W. contributed equally to this work.

\end{document}